\documentclass[12pt]{iopart}
\usepackage{graphicx}

\begin{document}

\title[Gingin seismic noise characterization]{Seismic noise characterization at a potential gravitational wave detector site in Australia}

\author{ Hamid Satari$^1$, Carl Blair$^1$, Li Ju$^1$, David Blair$^1$, Chunnong Zhao$^1$, Erdinc Saygin$^{1,2}$, Patrick Meyers$^{3,4}$ and David Lumley$^5$}

\address{$^1$ School of Physics, University of Western Australia, Australia}
\address{$^2$ CSIRO, Deep Earth Imaging Future Science Platform, Australia}
\address{$^3$ School of Physics, University of Melbourne, Australia}
\address{$^4$ Theoretical Astrophysics Group, California Institute of Technology, Pasadena, USA}
\address{$^5$ Dept Geosciences, Dept Physics, University of Texas at Dallas, USA}
\ead{hamid.satari@research.uwa.edu.au}
\vspace{10pt}
\begin{indented}
\item[]December 2022
\end{indented}

\begin{abstract}

A critical consideration in the design of next generation gravitational wave detectors is the understanding of the seismic environment that can introduce coherent and incoherent noise of seismic origin at different frequencies. We present detailed low-frequency ambient seismic noise characterization (0.1--10~Hz) at the Gingin site in Western Australia. Unlike the microseism band (0.06--1~Hz) for which the power shows strong correlations with nearby buoy measurements in the Indian Ocean, the seismic spectrum above 1~Hz is a complex superposition of wind induced seismic noise and anthropogenic seismic noise which can be characterized using beamforming to distinguish between the effects of coherent and incoherent wind induced seismic noise combined with temporal variations in the spatio-spectral properties of seismic noise. This also helps characterizing the anthropogenic seismic noise. We show that wind induced seismic noise can either increase or decrease the coherency of background seismic noise for wind speeds above 6~m/s due to the interaction of wind with various surface objects. In comparison to the seismic noise at the Virgo site, the secondary microseism (0.2~Hz) noise level is higher in Gingin, but the  seismic noise level between 1 and 10~Hz is lower due to the sparse population and absence of nearby road traffic.

\end{abstract}
\section{Introduction}

One of the objectives of next generation gravitational wave (GW) detection is to extend detector sensitivity to low frequencies down to 2~Hz \cite{punturo}. Detectors with a factor of 10 improvement in amplitude sensitivity aim to extend the horizon for blackhole coalescence events to encompass the entire Hubble volume \cite{punturo, helioseis}. A significant instrumental challenge to achieve these goals is to improve low-frequency performance, which requires detailed understanding of the seismic environment. Two factors are particularly important: a) Newtonian noise that cannot be shielded but can be modelled based on environmental measurements  \cite{saulson,hughes,creighton} and b) the wavelength and coherence of seismic noise that determine the range over which seismic signals are correlated \cite{low_coherency}. State of the art isolation systems are capable of suppressing seismic ground motion by 12 orders of magnitude above a few Hz \cite{12order}. However, challenges remain in improving the isolation systems for future detectors especially at low frequencies below 10~Hz partly because vibration isolation systems that enable detection of GWs above 10~Hz, have resonant frequencies around a few~Hz \cite{coughlin2014}. Suspension resonances are close to the microseism band where the ocean-earth coupled seismic wave energy peaks between 0.1--1~Hz \cite{Nishida2017}. This contributes to excess residual motion of suspended test masses and other optical components. Newtonian noise (NN) is predicted to be one of the limiting noise sources in third-generation GW detectors, such as the Einstein Telescope down to 2~Hz \cite{punturo, coughlin2014, ET_correlation}. It's suppression requires accurate characterisation of density fluctuations caused by seismic noise.

Seismic arrays have been widely proposed \cite{coughlin2014,Driggers, Coughlin_2016,Badaracco2019} to actively assist the vibration isolation systems by cancelling ground motion \cite{Abbott_2004}. The information obtained from a seismic array can be exploited by a feed-forward control system before the seismic vibration enters the system \cite{Feedforward,Derossa2012}. Arrays can also be used to measure the ambient seismic wavefield allowing prediction and subtraction of NN \cite{new_Virgo2021, Badaracco2019, Badaracco2020, Coughlin2016}. For an efficient isolation system, and for modelling the contribution of seismic noise to NN and its optimal subtraction, it is vital to classify seismic noise sources and their effects on the interferometer~\cite{LIGO_site, VIRGO_site, Virgo_2021}. Array processing methods like beamforming \cite{Rost2002} provide valuable information on the seismic wave-types and its modes which helps determining the relative strength of the noise source and the directivity of the seismic waves \cite{Koley_2022, directivity_ocean}.  

In this paper we show that beamforming can also be used to resolve site-dependant complexities because it separates the coherent and incoherent portions of the seismic wavefield and provides information about the seismic coherency over the array in the form of the variations in the array's beampower. In contrast to the microseism at the Gingin High Optical Power gravitational wave research Facility \cite{zhao} (HOPF) in Western Australia (commonly known as the Gingin site) where the single station noise power is strongly correlated with buoy measurements in the Indian Ocean, we cannot use the noise power for the seismic noise characterization above 1~Hz due to the complex superposition of wind induced and anthropogenic seismic noises. We use beamforming to monitor the temporal variations in the coherency, direction and wave speed of seismic noise to interpret the changes in the seismic noise power at the Gingin site. The Gingin site is one of the candidate sites for the proposed Neutron Star Extreme Matter Observatory (NEMO) in Australia \cite{NEMO_1}. This study will help us in understanding the characteristics of different seismic noise sources at different frequency bands for developing suitable vibration isolation and mitigation methods for GW detectors as well as assisting the site selection for future GW detectors such as a potential Cosmic Explorer in Australia (CE South) \cite{evans2021horizon, reitze2019cosmic}.

In section 2, we provide a general view on the properties of seismic noise at the Gingin site and compare them with those at the Virgo site in Italy that hosts the advanced Virgo GW  detector \cite{VIRGO_site}. Then we focus on the properties of the microseism, wind induced and anthropogenic seismic noises as the main low-frequency noise sources. In section 3, we characterize the primary and secondary microseism by comparing single station seismic noise power with oceanic wave data from two buoy stations in the Indian Ocean. In section 4, after a brief review of the f-k beamforming technique, we use it to analyze the temporal variations in the spatio-spectral properties of the ambient seismic noise. This analysis helps us to distinguish between the effects of coherent and incoherent wind induced seismic noise on the coherency of background anthropogenic seismic noise. In section 5, we characterize the physical properties of the subsurface by extracting the dispersion profile from the beamforming results for 0.1--10~Hz frequency band and discuss the uncertainty of the beamforming measurements. In section 6, we present our major conclusions and discuss the implications on GW detection.

\begin{figure}
\centering
\includegraphics{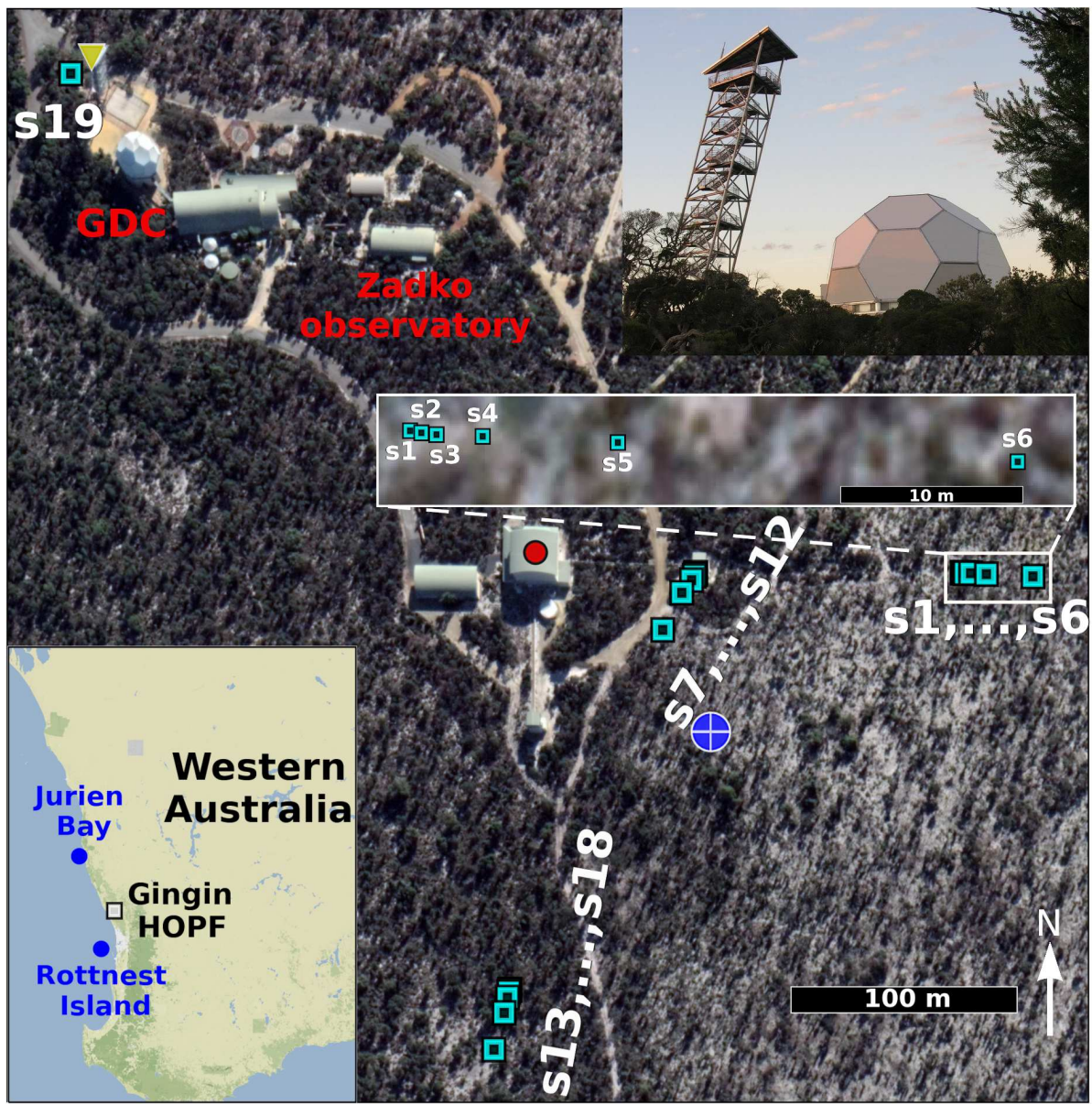}
\caption{Seismic array layout in the Gingin site on Google Earth image, Map Data @2022 Google.  18 seismometers (s1--s18) are used in three sub-arrays of six nodes with similar logarithmic spacing as shown in the zoomed inset. The blue dot with the white cross shows the geometric centre of the seismic array. An additional seismometer (s19) is deployed next to the leaning tower shown with the yellow triangle on its roof. Zadko observatory and GDC are located to the southeast of the tower. The inset on the top right of the figure shows the leaning tower and one of the GDC buildings. The red dot on the roof of the Central-Station shows the location of an anemometer used in this study. Blue dots in the left bottom inset map show the locations of Jurien bay and Rottnest Island buoy stations in the Indian ocean which are 141 kms and 67 kms away from the seismic array, respectively.}
\label{Figure_1}
\end{figure}

\section{General properties of Gingin ambient seismic noise}

The HOPF is located 20 km from Gingin town, 71 km north of the Perth city centre and 17 km east of the WA coast (figure~\ref{Figure_1}). There are three research lab buildings (the Central-Station, South End-Station and the East End-Station), housing two 80 m long suspended optical cavities. There is also a building for the Zadko observatory \cite{Zadko} as well as a public education centre, Gravity Discovery Centre (GDC) with several buildings and a 40~m tall leaning tower located northwest of the site (yellow triangle in the top left of figure~\ref{Figure_1}, also shown in the top right inset). Data used for this study were acquired during days 168--182 (July) of 2021 by an array of 19 broadband (T = 120 s) seismometers (3-component Nanometrics Trillium T120/PA), an anemometer (Intech WS3-WD-TB-CL) installed on the roof of the Central-Station, and two buoy stations in the Indian ocean (Oceanographic Services, WA Department of Transport) which are shown in figure~\ref{Figure_1}. 

\begin{figure}[t!]
\centering
\includegraphics{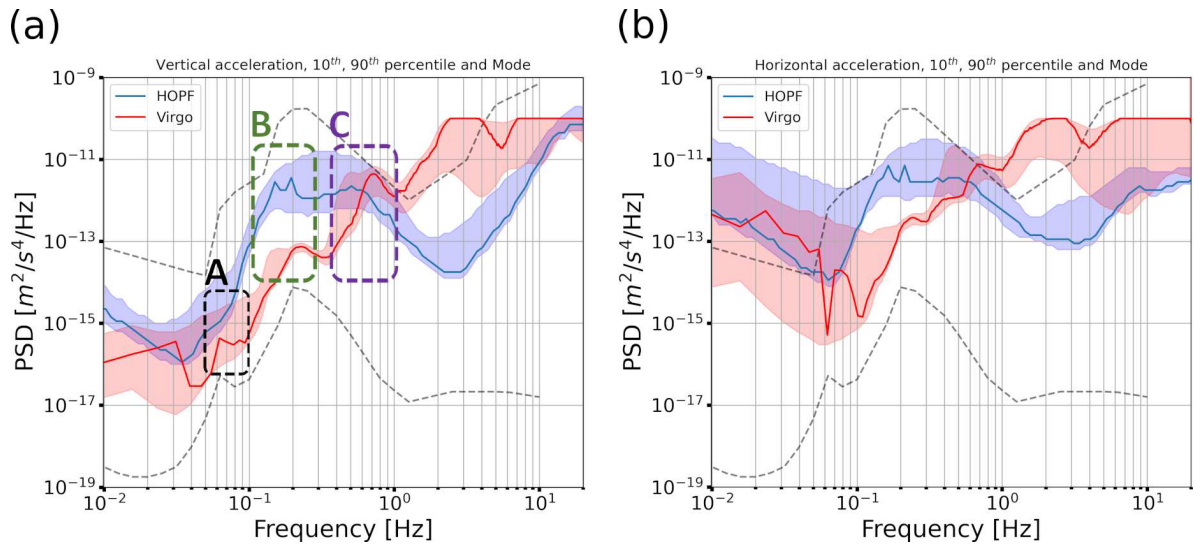}
\caption{Power spectrum of (a) the vertical and (b) the horizontal seismic noise measured by a broadband seismometer on the surface at the Gingin site (blue) compared to those recorded at virgo  (red) \cite{Koley_report}. These measurements were made during the same month between 20 June and 20 July 2019 at the two sites. The solid line shows the mode and the 10\textsuperscript{th}--90\textsuperscript{th} percentile is represented as the band. Three peaks associated to the primary, low-frequency and high-frequency secondary microseisms are highlighted respectively as A, B and C on the vertical component. In both graphs the dashed gray lines show the Peterson’s new low and high noise models (LNM/HNM).}
\label{Figure_2}
\end{figure}

The Gingin site is located in banksia bush land composed of shrubs and sparsely distributed trees. The nearest highway is 8~km away, and nearest town, Gingin, is 20~km away from the site with a population of only 900 (Australian Bureau of Statistics, 2021). In addition to the human activity during daytime inside the lab buildings, there are people working in or visiting the Zadko observatory and the GDC at days and nights including the weekends. There are a few farms sparsely located in the region. While the seismic environment at the Gingin site has many features common to other parts of the world, it has unique features in its spectrum that come from the local environment. To better clarify these features with a perspective on their relevance to GW detection, we compared the vertical and horizontal seismic probabilistic power spectral densities (PPSD) from a surface seismometer at the Gingin site with those at the Virgo site \cite{Koley_report} during the same time between 20 June and 20 July 2019 (figure~\ref{Figure_2}). Peterson’s new high- and low-noise models \cite{peterson} are plotted for comparing the seismic noise levels with the global average.

The primary (0.06--0.1 Hz) and secondary (0.1--1~Hz) microseisms \cite{Nishida2017, gal_2015, Hasselmann_1963, Beker_2015} are highlighted with three peaks labeled as A, B and C in the vertical PPSD plot. Primary microseism (marked by A) is caused by the interaction of pressure fluctuations from infragravity ocean waves with ocean floor in the coastal areas \cite{gal_2015, Hasselmann_1963}. At some locations, there are two peaks associated with the secondary microseism. The low-frequency secondary microseism (0.1--0.3~Hz, marked by B) is made by standing waves with double frequencies as those of the primary microseism generated through the interaction of ocean swells with their reflections from the ocean floor \cite{gal_2015, WavewatchIII, Hasselmann_1963, Gerstoft_2006}. The third peak, referred to as high-frequency secondary microseism (0.3--1~Hz, marked by C), is caused by local sea waves generated by local sea wind \cite{gal_2015}. At the Virgo site it is believed to be caused by fluctuations in the wave heights in the Mediterranean sea \cite{Beker_2015}. The source of the primary and low-frequency secondary microseisms (A and B) is often considerably far from the coasts and can create complex superposition of seismic body waves and surface waves \cite{gal_2015, Coughlin_2019}. Comparing the general properties between the two sites shows that the primary and high-frequency secondary microseisms (A and C) have almost the same strength, while the low-frequency secondary microseism (B) is $\sim$ two orders of magnitude lower at Virgo and the seismic noise above 1 Hz is $\sim$ two orders of magnitude lower at Gingin. At the Virgo site, the microseism is generated by the northern Atlantic ocean, and at Gingin it comes from the Indian ocean. The storm systems, seasonal peaks, and coupling mechanisms are thus different for the two sites that lead to the differences between their PPSDs.

\begin{figure}
    \centering
    \includegraphics{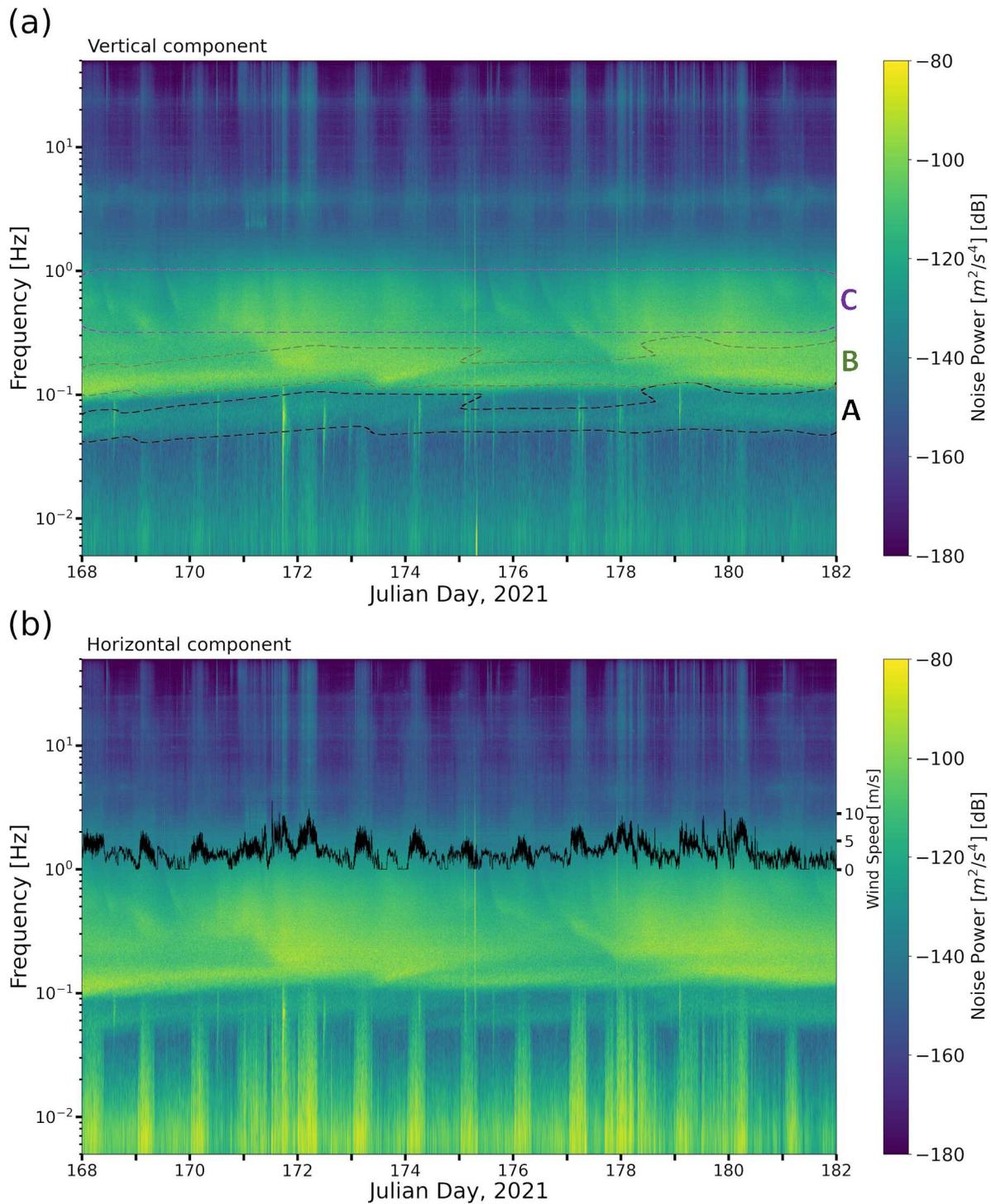}
    \caption{Spectrogram of (a) the vertical and (b) the horizontal component of seismometer s12 next to the East End-Station (figure~\ref{Figure_1}). The data was recorded during days 168--182 of 2021. Each day in the figure starts at 8~am local time. Three peaks associated to the primary, low-frequency and high-frequency secondary microseism are highlighted respectively as A, B and C on the vertical component similar to those in figure\ref{Figure_2}(a). Scaled wind speeds (black) are overlaid on the horizontal component. Wind induced seismic noise is evident in the form of a diurnal variation of the power correlated with wind speeds across much of the spectrum especially below 0.1~Hz on the horizontal component.}
    \label{Figure_3}
\end{figure}

One way to identify and characterize the noise sources above 1~Hz is to examine the spectrum over time. We use spectrograms of the vertical and horizontal surface seismic data acquired during two weeks in July 2021 at location s12 (figure~\ref{Figure_1}). The vertical component spectrogram shown in figure~\ref{Figure_3}(a) better illustrates the underlying dominant seismic noise e.g. the microseism peaks around 0.08~Hz, 0.15~Hz and 0.6~Hz highlighted by A, B and C, respectively. An interesting feature is the similar temporal variation of the low-frequency secondary microseism (B) to that of the primary microseism (A) but with double frequencies. On the other hand, the horizontal component spectrogram (figure~\ref{Figure_3}(b)) more clearly shows the daily power increase in the first half of the Julian days, equating to 8~am to 8~pm local time. In figure~\ref{Figure_3}(b), we have overlaid wind speed measurements in black (the scale is on the right side of the figure). 

It is clear that an increase in wind speed corresponds to an increase in the seismic noise power below 0.1 Hz and above 1 Hz. This is the wind induced seismic noise with a daytime periodicity that has dominated other seismic noise in several frequency bands in the background. The primary microseism peak around 0.08~Hz, marked by A in figure~\ref{Figure_3}(a) is an example that can be distinguished from the wind induced seismic noise because it is better visible in the vertical component while it is partially masked in the horizontal component. In contrast, it is difficult to characterize anthropogenic seismic noise above 2~Hz by a common daily variation in the seismic power due to the presence of wind induced seismic noise with a diurnal pattern at both components.

Comparing figures~\ref{Figure_2} and~\ref{Figure_3} with the observed daily seismic RMS changes associated with human activity at the Virgo site \cite{VIRGO_site}, reveals the main reason for the differences between the two sites. Weaker anthropogenic seismic noise is common in remote rural areas \cite{USA_3Sites} like Gingin where other sources, like wind induced seismic noise, become more apparent. This explains the three orders of magnitude lower seismic noise power for the Gingin site in the 2--4~Hz frequency band than that for the Virgo site  (figures~\ref{Figure_2}). The absence of highways in the vicinity of the Gingin site means that seismic noise generated by local traffic is significantly lower than that at the Virgo site in this frequency band \cite{Virgo_2017}. Likewise, an order of magnitude lower power from 6~Hz to 10~Hz at the Gingin site compared to that at the Virgo site is explained by sparser population nearby \cite{VIRGO_site, Beker_2012}. 

\section {Microseism}

There are 19 seismometers in the array with minimum 0.5~m (s1--s2, s7--s8 and s13--s14) and maximum 540~m (s6--s19 and s13--s19) sensor separations (figure~\ref{Figure_1}). Depending on the seismic velocity structure of the site, these values set a maximum and a minimum threshold respectively for the frequencies of the incident wavefield of which properties can be reliably measured by the array \cite{Foti2017, Wathelet2008, Coughlin_2019}. The configuration of this seismic array has been originally designed to study low coherency of wind induced seismic noise in three distinct locations \cite{low_coherency} and it cannot be used for beamforming analysis of microseism. Its aperture size (540~m) is far shorter than the wavelength of the microseism ($\sim$ 30~km at 0.1~Hz) which leads to uncertain velocity measurements (discussed in section 5). Therefore, here we focus only on the single station seismic data. 

\begin{figure}
    \centering
    \includegraphics{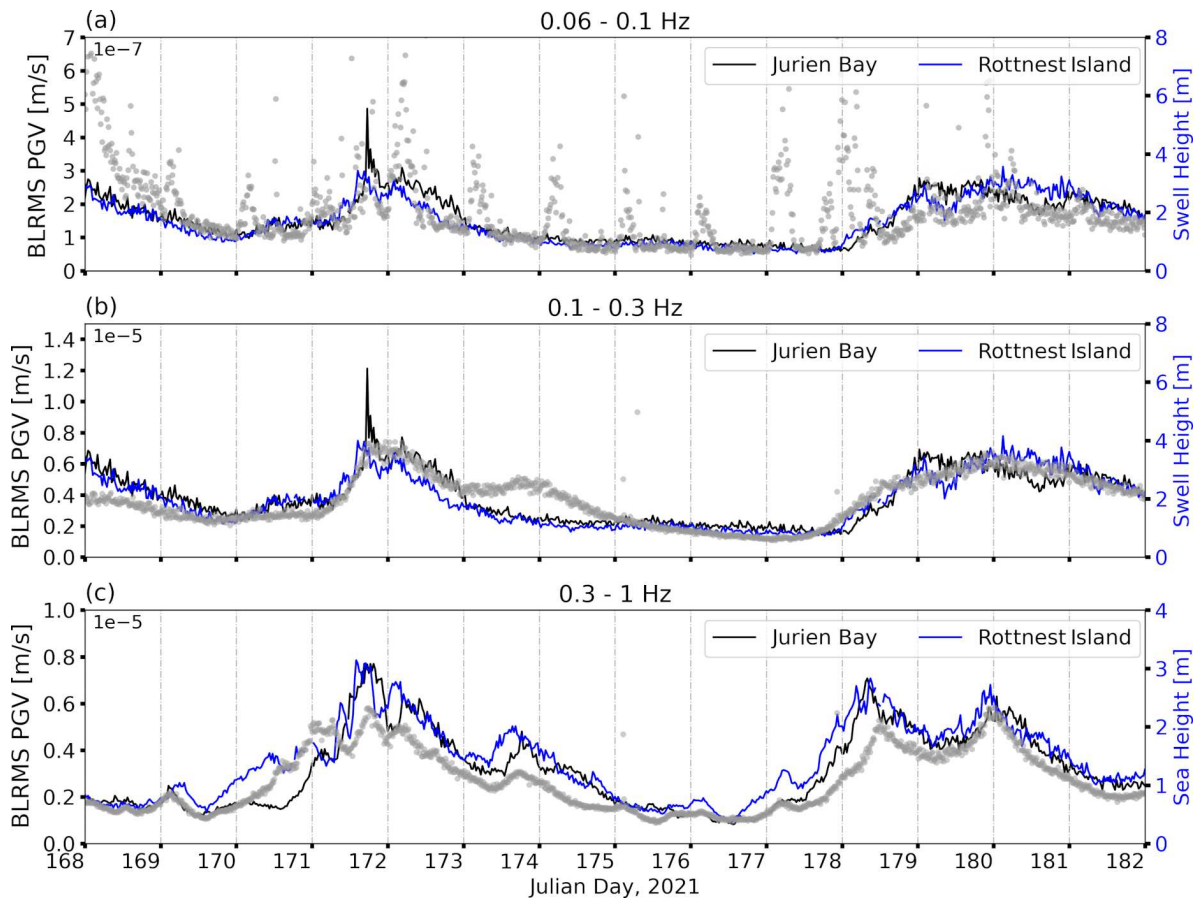}
    \caption{Temporal variations in the bandlimited RMS peak ground velocity (BLRMS PGV) computed using 15 minutes non-overlapping intervals for (a) primary microseism (0.06--0.1~Hz) which is correlated with swell wave heights except for the windy times with higher RMS peak ground velocities during days (each day starts at 8~am local time), (b) low-frequency secondary microseism (0.1--0.3~Hz) which is correlated with swell wave heights and (c) high-frequency secondary microseism (0.3--1~Hz) which is correlated with sea wave heights. The wave data were measured at Jurien Bay and Rottnest Island buoy stations in the Indian ocean, located 141 kms and 67 kms to the west of the seismic array, respectively.}
    \label{Figure_4}
\end{figure}

Figure~\ref{Figure_4} shows the correlations between the seismic RMS peak ground velocity and the wave data measured at Jurien Bay and Rottnest Island buoy stations in the Indian ocean which are located 141 kms and 67 kms to the west of the seismic array, respectively. We used 15 minutes non-overlapping intervals of the band-pass filtered vertical seismic data over the two weeks. Figure~\ref{Figure_4}(a) and (b) show that the primary microseism (0.06--0.1~Hz) and the low-frequency secondary microseism (0.1--0.3~Hz) are both correlated with swell wave heights which implies that they originate from the same source. The exception to this is the primary microseism with higher seismic RMS peak ground velocity during day times which is a local effect due to wind induced seismic noise (figure~\ref{Figure_3}). We have previously observed that the wind induced seismic noise at this frequency band is incoherent \cite{low_coherency}. Figure~\ref{Figure_4}(c) shows the high-frequency secondary microseism (0.3--1~Hz) which is correlated with sea wave heights. The different temporal variations of the high frequency secondary microseism power is representative of a separate noise source (local sea waves hitting the WA shoreline) from that of the primary and low-frequency secondary microseisms (distant swell waves). 

The correlation of the seismic data with two buoy stations more than 200~kms apart from one another shows that the microseismic wave energy mainly comes from the west. Increased swell heights on days 171, 172 and between days 179 and 182 (figures~\ref{Figure_4}(a) and (b)) is suggestive of storms in the Indian ocean (distant source) which have increased the ocean-Earth coupled microseism energy. Such variations have been reported to be accompanied by seismic wave-type transition from body waves to surface waves and vice versa as it was explained by \cite{Coughlin_2019, Pat_2021, Gerstoft_2006} for secondary microseism at 0.2~Hz. Surface waves associated with oceanic storms dominate the lower amplitude body waves, while in the absence of storms, body waves can be detected by lower amplitude and higher coherency. This results in an anti-correlation between the seismic noise power and seismic coherency which can be monitored along with the corresponding changes in the azimuths \cite{directivity_ocean} and wave-speeds\cite{Coughlin_2019} using a seismic array with a configuration optimized for this purpose \cite{seismic_arrays}.

\section {Monitoring seismic spatio-spectral properties by beamforming}

The general idea of beamforming in seismology is to amplify the coherent portion of seismic wave energy and suppress the incoherent noise \cite{Krim1996TwoDO, Rost2002, Gal}. As an array processing method, it enables monitoring the temporal variation of the spatio-spectral properties of the seismic wavefield. This results in a useful separation between incoherent and coherent wind induced seismic noise and between incoherent wind induced seismic noise and coherent anthropogenic seismic noise at the Gingin site. Applying beamforming to GW seismic noise site characterization is gaining attention \cite{Virgo_2021, Koley_2022, Coughlin_2019, Virgo_2017}. Here we emphasize on its efficiency in distinguishing between various noise sources and different wave-types based on their different coherency states.

We use frequency-wavenumber (f-k) beamforming \cite{Kelly1967} that maximizes the array beam power in the frequency domain. The frequency-domain signals of different sensors are phase delayed and stacked. This method assumes a 2D Earth model with spatial coordinate vector \textbf{r} and a coherent plane wave traveling with horizontal velocity $v$ ($km/s$), then it simultaneously calculates the azimuth $\theta$ and slowness $s=v^{-1}$ ($s/km$) of the seismic wavefield \cite{smart1971, capon1973}. The coherent portion of the seismic surface waves will be recorded at different locations $\textbf{r}_i$ of the array with corresponding travel times depending on the subsurface seismic velocity structure and the distances of the seismometers from the centre of the array. The phase delays for all stations can be calculated in the form of a steering vector \cite{Gal}

\begin{equation}
    \textbf{a}(f,\textbf{s}) = e^{-2\pi{if}\textbf{s}(\textbf{r}_{n}-\textbf{r}_{0})},    
    \label{eq1}
\end{equation}

with

\begin{equation}
    \textbf{s}= (s \sin\theta,s\cos\theta)
    \label{eq2}
\end{equation}in which $\textbf{r}_{n}-\textbf{r}_{0}$ is the distance vector between $n$\textsuperscript{th} sensor and the reference sensor located at the centre of the array $\textbf{r}_{0}$. The azimuth ($\theta$) is measured clockwise from north and shows the directions from which the seismic wavefield arrived at the array. Then the beam power associated with the azimuth and magnitude of $\mathbf{s}$ is calculated by applying the phase delays to the cross-spectral density of the Fourier transformed seismic traces $\textbf{X}(f)$:

\begin{equation}
    P (f,\textbf{s}) =  \frac{1}{L^2M^2} \textbf{a}^H(f,\textbf{s})\textbf{X}(f)\textbf{X}^H(f)\textbf{a}(f,\textbf{s}),
     \label{eq3}
\end{equation} where $L$ and $M$ are respectively the window length and the number of the seismometers, and $^H$ denotes the Hermitian transpose. $\textbf{X}(f)\textbf{X}^H(f)$ is the Hermitian cross spectral density matrix (CSDM) that stores all of the auto-spectral densities in its diagonal terms and cross-spectral densities in its off-diagonal terms. By using a grid search that iterates over the slowness and azimuth, the beamformer (eq.~(\ref{eq3})) estimates the power contributions from a variety of synthetic plane waves. The phase shifts generated by matching slowness and azimuth to those of the recorded plane wave stored in the CSDM will result in coherent phase stacking and maximum beam power at a certain frequency bin. By repeating this analysis over consecutive windows of the recorded seismic traces, temporal variations in the maximum beam power and the corresponding azimuths and slownesses are calculated. This provides additional information on the properties of the seismic wavefield. 

Theoretically, the normalization factor $\frac{1}{L^2M^2}$ in eq.~(\ref{eq3}) means that the beam power must be 1 for a signal that is perfectly coherent across the array. However, in practise, there is always contribution of incoherent noise from different azimuths that prevents the array to be ideally tuned on a specific noise source \cite{seismic_arrays}. The higher the coherency, the beam power will be closer to 1 which can be used to infer the relative variations in the seismic coherency at a specific frequency band over time. 

\subsection{Wind induced seismic noise}

Spatial and temporal fluctuations in the atmospheric pressure loads on the Earth’s surface, create local ground tilt \cite{Sorrells, Peterson_1976}. In a recent study in the Gingin site \cite{low_coherency} we showed that wind induced seismic noise affects the seismic coherency in two ways depending on its interaction with different surface objects at different frequencies. First, it is incoherent and masks the background coherent seismic noise during local daytime e.g. the primary microseism (0.06--0.1~Hz), etc. Second, when it excites resonant features in the landscape, it injects coherent seismic noise (several spectral lines between 4 to 9~Hz). Because of the spread of the wind induced seismic noise across the large portion of the seismic spectrum in the site, we have previously studied its low coherency via a dedicated coherence length analysis and discussed its implications for advanced GW detection \cite{low_coherency}. We analyse this effect using beamforming in the next sub-section.

\begin{figure}
    \centering
    \includegraphics{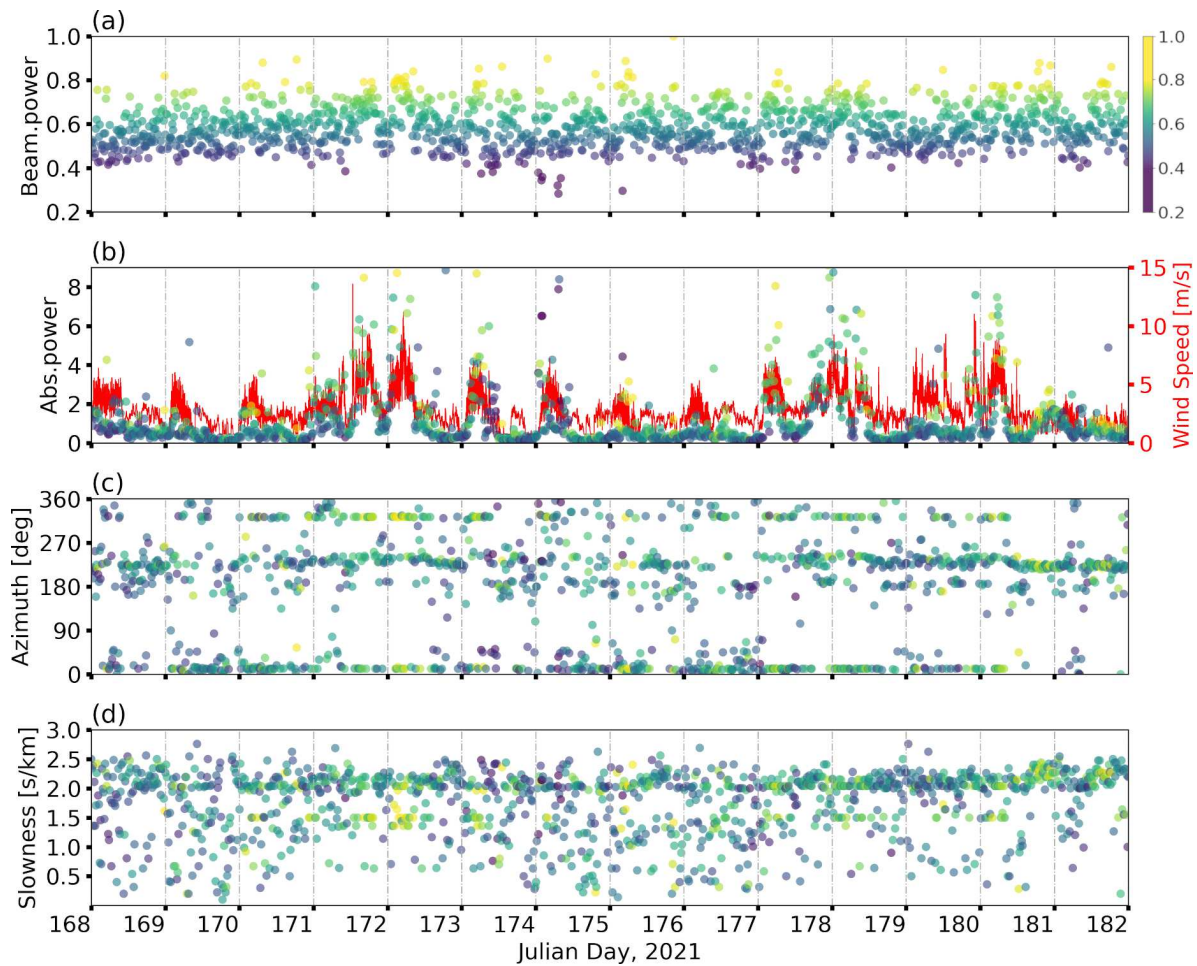}
    \caption{Beamforming results for vertical seismic noise data band-pass filtered between 4.2--4.3~Hz acquired during days 168--182 of 2021. (a) Variation of the beam power of which colors are used as the reference for the other measurements, (b) the absolute power of the reference seismometer in the centre of the array which is compared against the wind speeds, (c) the azimuths and (d) the slownesses that resulted in the maximum beam powers. High beam powers (bright green and yellow) occur for wind speeds above 6~m/s and show the coherent wind induced seismic noise with convergent azimuth around 320 degree separated from the other features with lower coherency and absolute power (dark green and blue).}
    \label{Figure_5}
\end{figure}

\begin{figure}
    \centering
    \includegraphics{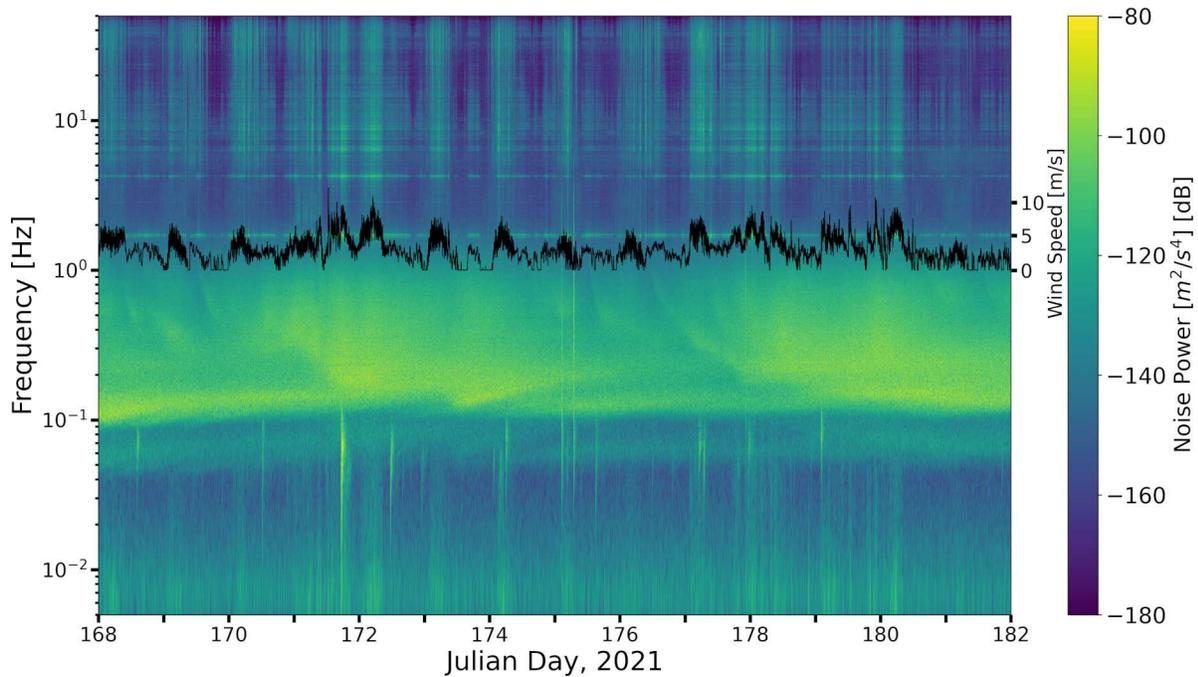}
    \caption{Spectrogram of the vertical component of seismometer s19 next to the leaning tower (figure~\ref{Figure_1}). Scaled wind speeds (black) are overlaid for visual correlation of the vibration modes of the tower with wind from 4.2~ Hz to 9~Hz. There is another spectral line at 1.8~Hz which is different form the wind-driven vibration modes.}
    \label{Figure_6}
\end{figure}

Investigation on the coherent version of the wind induced seismic noise in the Gingin site revealed that there are several wind-driven coherent spectral lines between 4--9~Hz. Figure~\ref{Figure_5} shows the beamforming results applied on the most coherent spectral line between 4.2--4.3~Hz. We band-pass filtered vertical component seismic data, and used 200 seconds long non-overlapping windows. Seismometer s19 next to the leaning tower (Figure~\ref{Figure_1}) was excluded from this analysis to guarantee enough source-array distance assumed for the beamforming algorithm regarding the leaning tower being a candidate source for the vibrations.

Figure~\ref{Figure_5}(a) shows the array beam power ($P (f,\textbf{s})$ in eq.~(\ref{eq3})) along the azimuths that have maximized it. As it can be seen from the color-bar, the higher the beam power, the brighter the colors. Therefore, the colors of the other three sub-plots (figure~\ref{Figure_5}(b)--(d)) are set based on their correspondent beam powers to help the reader follow the impact of the changes in the coherency of the wavefield on the other instantaneous measurements. Figure~\ref{Figure_5}(b) shows the absolute power of the reference seismometer ($X_{0}(f)X^H_{0}(f)$) at the geometric centre of the array. The absolute power of the rest of the seismometers which have not been shown here are synchronised to that of the reference seismometer using the corresponding phase shifts. This in turn results in the synchronization of the beam power with the absolute power. 

The magnitude of the absolute power is proportional to the window length. Here, we are only interested in its temporal variation and correlation with the wind speeds.  Values with bright colour (bright green and yellow) in all of the plots show that the measurements made at the times with wind speeds above 6~m/s have high coherency (because of their high beam power). The coherent portion of the wavefield which is correlated with wind and has convergent azimuth values (bright colors clustered around 320 degree) is separated from the less coherent features (dark green and blue dots). The azimuth ($\theta$) in figures~\ref{Figure_5}(c) shows the directions from which the seismic wavefield with maximum beam power (figure~\ref{Figure_5}(a)) arrived at the array. The azimuth at the times with wind speeds above 6~m/s (320 degree) is in the north-westerly direction consistent with the location of the tower (40~m tall) shown in figure~\ref{Figure_1}. The slowness amplitudes ($s$) along these azimuths are represented in figures~\ref{Figure_5}(d). These values are about 1.6 s/km i.e. wave speeds are about 625 m/s which is consistent with surface wave propagation at this frequency. Two other sources are present around 220 and 20 degrees with slowness values mainly between 2 and 2.5 s/km i.e. wave speeds between 400 and 500 m/s. 

The spectrogram shown in figure~\ref{Figure_6} is calculated from the vertical seismic data measured by seismometer s19 next to the leaning tower. There are several spectral lines between 4--9~Hz including one around 4.2~Hz. These lines do not exist in the spectrogram shown in figure~\ref{Figure_3} for seismometer s12 which is 400~m away from the tower. For wind speeds above 6~m/s, the power of these lines increases with wind speed and makes them the dominant feature in the spectrogram. This supports the beamforming results for the coherent wind induced seismic noise and determines the leaning tower with several wind-driven vibration modes as the source for this anomaly.

\subsection{anthropogenic seismic noise}

\begin{figure}
    \centering
    \includegraphics{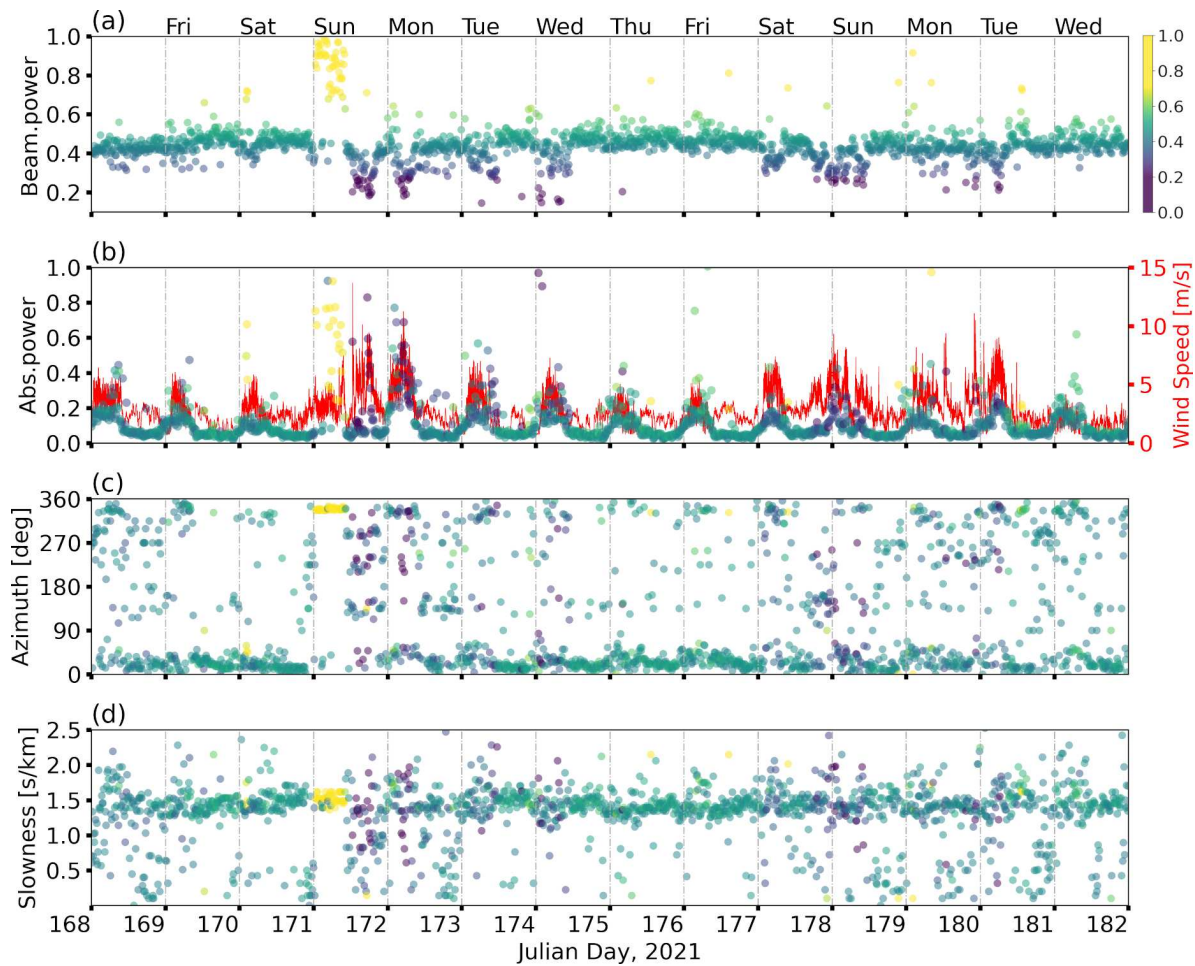}
    \caption{Beamforming results for vertical seismic noise data band-pass filtered between 2--4~Hz acquired during days 168--182 of 2021. (a) Variation of the beam power of which colors are used as the reference for the other measurements, (b) the absolute power of the reference seismometer in the centre of the array which is compared against the wind speeds, (c) the azimuths and (d) the slownesses that resulted in the maximum beam powers. Low beam powers (dark blue) occur for wind speeds above 6~m/s and show the incoherent wind induced seismic noise with smeared azimuth and slowness values. A ten hours long coherent anthropogenic seismic noise has sharply increased the beam power on day 171 from 8~am to 6~pm (yellow) local time with a narrow azimuth range pointing towards the north-west of the site (330 degree).}
    \label{Figure_7}
\end{figure}

Single station seismic noise power (figure~\ref{Figure_3}) shows that the seismic spectrum is dominated by wind induced seismic noise above 1~Hz. Monitoring the seismic spatio-spectral properties helps to gain more information about the background anthropogenic seismic noise. Figure~\ref{Figure_7} shows the beamforming results for 2--4~Hz using 250 seconds long non-overlapping windows. Opposite to the coherent wind induced seismic noise (figure~\ref{Figure_5}), there is an anti-correlation between the beam power and wind speeds higher than 6~m/s (dark blue values) which shows that wind induced seismic noise is incoherent at this frequency band. Overall, the beam power is constantly low (an average of 0.4) and the absolute power correlates with wind speed except for day 171 from 8~am to 6~pm local time (yellow), with 330 degree azimuths which could be due to a maintenance or a special event on the GDC which is open on weekends. A road at the north-west outside the site can also be the candidate source. This shows that incoherent wind induced seismic noise is dominant except at times when the power and coherency of anthropogenic seismic noise increase due to a temporary event.

The seismic PSD starts to increase from 3~Hz for the vertical and from 4.5~Hz for the horizontal component (figure~\ref{Figure_2}) up to 10 Hz. Figures~\ref{Figure_5} to~\ref{Figure_7} show that the background seismic noise in this frequency band is dominated by incoherent wind induced seismic noise while there are also several coherent wind-driven spectral lines generated by the leaning tower. This means that several frequency bins between the spectral lines are required to separate between the effects of coherent and incoherent wind induced seismic noise on the anthropogenic seismic noise for 4.3--10~Hz frequency band. Furthermore, because the array is not concentric, the short sensor separations in the array are in distinct locations in the three sub arrays (s1--s18 in figure~\ref{Figure_1}). These factors, make it inefficient to use beamforming for monitoring the spatio-spectral properties of the anthropogenic seismic noise at higher frequencies. Instead, in the next section, we provide a statistical perspective on the rest of the target frequency band.

\section{Discussion}

The slowness resolution of an array is controlled by its aperture size, sensor density and configuration \cite{seismic_arrays, Gal}. The f-k beamforming under the assumption of plane wave propagation, uses data independent weights (steering vector, eq.~(\ref{eq1})) for the seismic traces in the summation process. These weights are solely controlled by the geometry of the array \cite{Gal}. As a result of this and because the seismic wavefield is sparsely sampled in the space domain, the array's capability to resolve slownesses in the f-k domain is directly subject to its geometry \cite{Wathelet2008}. We cannot analyze the temporal variation in the directivity and wave speed of the microseism using the current array due to its limited aperture size.

\begin{figure}
    \centering
    \includegraphics{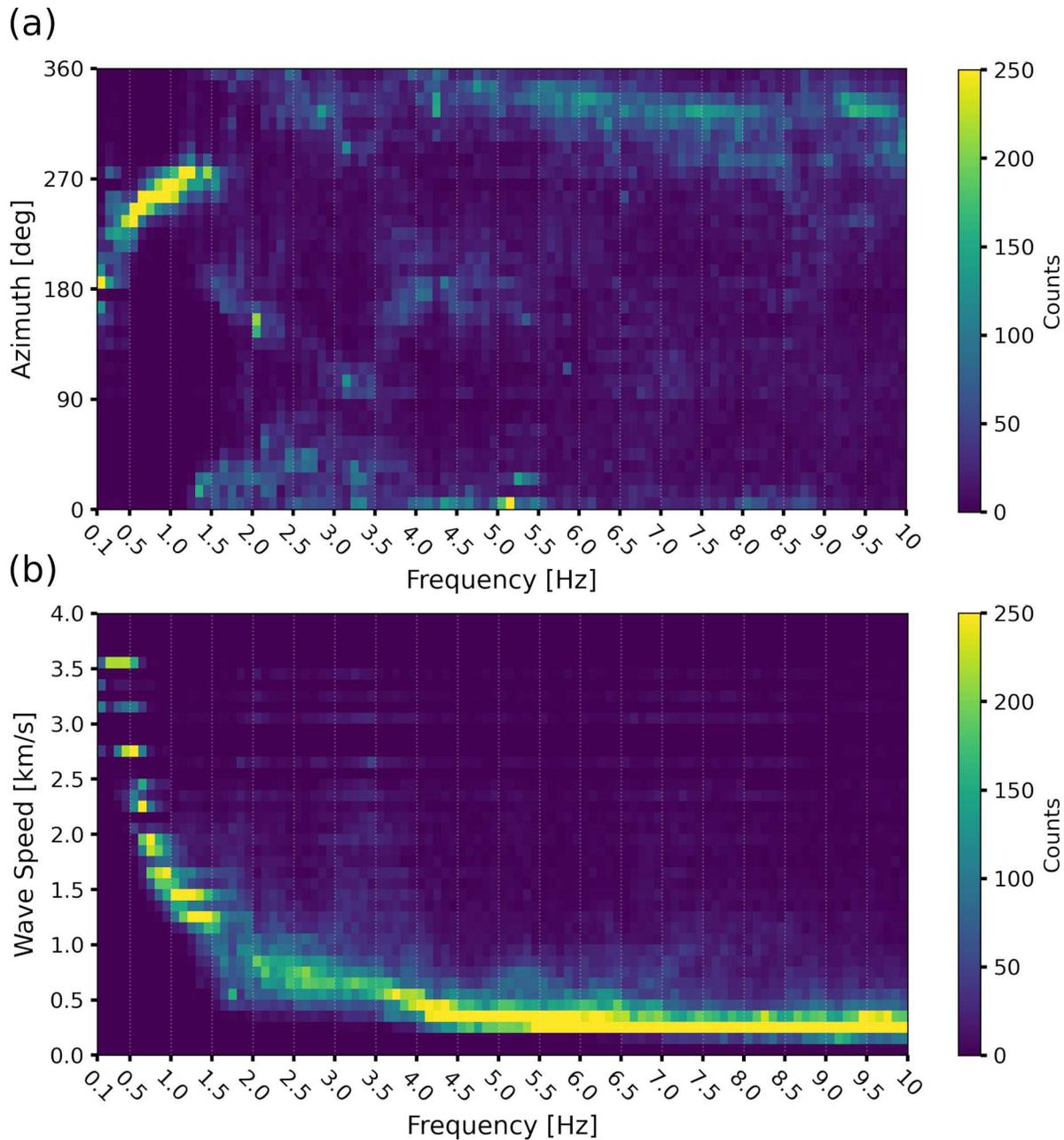}
    \caption{Histograms of (a) azimuth and (b) wave speed (dispersion plot) as a function of frequency for 0.1--10~Hz with 0.1~Hz frequency steps. The histograms are obtained from 79200 beamforming measurements made by the data from 19 seismometers acquired during days 172--182 of 2021.}
    \label{Figure_8}
\end{figure}

By monitoring the temporal variations in the power contributions from various azimuths and slownesses (spatio-spectral properties), we characterized the noise sources between 2 and 4.3~Hz using beamforming. In order to provide an estimate on the uncertainty of those measurements and to extend the analysis statistically to lower and higher frequencies, we calculate beamforming in the 0.1--10~Hz frequency band using finer frequency bins. Figure~\ref{Figure_8} shows the azimuth vs frequency and the wave speed vs frequency (dispersion plot) obtained from beamforming results using 0.1~Hz frequency bins for Julian days 172-182 of 2021. Each line in these plots is a histogram of the beamforming measurements at the frequency bin and the brighter colours mean higher histogram counts. 

The histogram of azimuth vs frequency in figure~\ref{Figure_8}(a) shows that the azimuth of the microseism ranges between $\sim$160 and $\sim$280 degrees and that the azimuthal uncertainty increases significantly below 0.6~Hz. On the other hand, the histograms of wave speed are multi-modal below 1.5~Hz with large standard deviations that increase towards 0.1~Hz. This happens because the wavefronts are so larger than the aperture of the array that they arrive at all of the stations almost simultaneously. This results in very large apparent wave speeds, hence very small slownesses below the slowness resolution of the array. The grid search finds high beam powers with quantized slowness values that in turn results in the multi-modal, uncertain and large wave speeds in the dispersion histograms below 1.5~Hz. 

The beamforming results are reliable above 1.5~Hz. The seismic noise in the band between 1.5 and 5.5~Hz is multi-directional with no preferred direction. This could be the result of the local machinery, human and farm activities. Above 5.5~Hz, the noise mainly comes from north and north-west most likely due to the human activities in the Central-Station and GDC buildings inside the site, or a road at the north-west outside the site. The dispersion plot carries information about the geology of the study area beneath the array. This information is essential for modelling Newtonian Noise in gravitational wave detector sites \cite{Koley_2022, new_Virgo2021}. First it shows that the contribution of seismic surface waves on the surface seismic wavefield is dominant and that there are minimal complexity like reflections and refractions in the study area, as each of these features would appear as an additional curve in the dispersion plot. Besides the fundamental mode of the surface waves with the dominant counts (bright green), there may be higher modes with weaker contributions around 2~Hz, between 3 to 4~Hz and between 6.5 to 8.5~Hz which could not be clearly resolved due to the limited spatial resolution of the current array. The presence of such overtones at different frequencies could be suggestive of soft soil to hard rock transition at different depths \cite{Foti2017, Koley_2022}. Detailed geological information (depth and shear wave velocity) can be derived from the inversion of the dispersion plot which is a non-unique problem.

\section{Conclusions and future works}

A combination of seismic, wind and ocean buoy measurements have allowed us to characterise the general properties of seismic noise at the Gingin site, and compare them to those at the Virgo site from 0.01 to 20~Hz. We also used a seismic array composed of 19 seismometers and the f-k beamforming technique for a detailed seismic noise characterization at the Gingin site between 1 and 10~Hz. We have shown that monitoring the variations in the seismic coherency over an array in the form of the variations in the array's beam power along with other spatio-spectral properties of the seismic noise, all obtained using beamforming, helps to better interpret the variations in the single station seismic noise power. This method can be applied to characterization of other sites with their unique seismic noise profiles and complexities.

Comparing the broad spectral characteristics of seismic noise between the Gingin and the Virgo site showed that the Gingin site has 1--3 orders of magnitude less seismic noise power above 1~Hz, but two orders of magnitude larger power in the 0.1--0.3 Hz band. Results were obtained using simultaneous data from both sites (July 2019). While it is possible that the differences relate to the opposite seasonality in the southern hemisphere, we have observed broadly similar results in the 0.1--0.3~Hz band for one month of summer data at the Gingin site. Long term data analysis is needed for more complete understanding of the seasonal variation of the microseism. 

The anthropogenic seismic noise at the Gingin site is between one to three orders of magnitude weaker for 1--10~Hz compared to those at the Virgo site. Together with the information obtained from the vertical and horizontal spectrograms, these features show that the Gingin site is located in a seismically rural area where wind induced seismic noise dominates the anthropogenic seismic noise with a diurnal variation in the seismic noise power. We have previously studied the incoherent wind induced seismic noise. The very short coherence length means that wind induced seismic noise is a local effect that has a broad-band signature as shown in the spectrograms. This makes it a challenging noise component for gravitational wave observatories, because it cannot be predicted using sparse spatial sampling. This means that isolation against wind induced seismic noise must make use of seismic measurements within short distances comparable to the coherence lengths. Primary microseism (0.06--0.1~Hz) is another noise component of which coherency and thus predictability using a seismic array is limited by the incoherent wind induced seismic noise. We note that the microseismic band is below the target frequency band of terrestrial gravitational wave detectors and its importance relates to the design of the control systems.

As previously observed, not all of the seismic noise above 1~Hz is incoherent. Single station data analysis cannot resolve the complex superposition of wind correlated seismic noise and anthropogenic seismic noise with the diurnal variation. We used f-k beamforming to monitor the temporal variations in the array's beam power that helped us to distinguish between coherent and incoherent wind induced seismic noise and their effects on the anthropogenic seismic noise. We found the most coherent vibration mode of the 40~m tall tower between 4.2 and 4.3~Hz among the other spectral lines present in the site from 4--9~Hz for wind speeds above 6~m/s. Beamforming results also showed how the wind induced seismic noise in the 2--4~Hz frequency band reduces coherency of anthropogenic seismic noise during times with wind speeds above 6~m/s. The anthropogenic seismic noise in this frequency band occasionally dominates the incoherent wind induced seismic noise with increased coherency and strength. 

Beamforming allowed estimation of wave direction and velocity as a function of frequency from 0.1--10~Hz showing the strong frequency dependence of wave speeds. This also provided information about the spatial distribution of the noise sources in the study area. The velocity histogram showed that wave speeds estimates are uncertain below $\sim$1.5~Hz due to the limited aperture size of the array, but otherwise the dispersion plot is reliable. This analysis showed that surface waves mainly contribute to the ambient seismic noise wavefield while there are some evidence of overtones in few bands probably because of different elastic moduli of subsurface layers at various depths. The azimuth histogram showed that the microseism come from the west (as expected), but there is no preferred direction of the seismic field from 1.5~Hz to 5.5~Hz. From 5.5~Hz to 10~Hz, the seismic waves mainly come from the north and northwest of the array most likely as a result of human activities within the site and/ or the northwest road outside the site. The results of this paper will be used to develop a data-driven seismic array configuration optimization method by which we will design a low-frequency seismic array needed for subtraction isolation proposed for gravitational wave detectors. We will also invert the dispersion curve and obtain geology information required for NN modelling.

\section{Acknowledgement}

This project was supported by Australian Research Council (ARC) Centre of Excellence for Gravitational Wave Discovery (CE170100004), ARC LIEF grant (LE200100008) and CSIRO Deep Earth Imaging Future Science Platform. The authors would like to also thank Siobain Mulligan and the Oceanographic Services at the Department of Transport of western Australia for providing the Indian Ocean wave data (buoy measurements) that was used in this study. Finally, we appreciate Dr. Soumen Koley at Gran Sasso Science Institute in Italy for providing us with the seismic data of the Virgo site that was used in this paper.  

\section{References}



\providecommand{\newblock}{}

\end{document}